\documentclass{elsart}

\usepackage{amssymb}
\usepackage{amsmath}
\usepackage{array}
\usepackage{pstricks}
\usepackage{pst-node}
\usepackage{graphicx}
\usepackage{pst-eucl}

\begin{document}

\begin{frontmatter}


\title{The relativistic velocity addition law optimizes a forecast
gambler's profit}
\author[labela1]{Edward W.~Piotrowski}
\author[labela2]{, Jerzy Luczka}
 \address[labela1]{Institute
of Mathematics, University of Bialystok, Lipowa 41, Pl-15424 Bialystok,
Poland; ep@alpha.uwb.edu.pl
}
\address[labela2]{Institute
of Physics, University of Silesia, Uniwersytecka 4, Pl-40007 Katowice,
Poland; luczka@us.edu.pl
}



\begin{abstract}
We extend the projective covariant bookmaker's bets model to the forecasting gamblers
case. The probability of correctness of forecasts  shifts
probabilities of branching. 
The formula for the shift of probabilities leads to the velocity
addition rule of the  special theory of relativity. In the absence of information about 
bookmaker's wagers the stochastic logarithmic rates completely determines the optimal stakes of forecast gambler.
\end{abstract}

\end{frontmatter}

\section{Introduction}
First, the true in Aristotle's law of syllogism, the individuals' preferences, the market prices are transitive \cite{piosla}. The transitivity is the fundamental property of effective markets. (However the different point of view was analysed e.g\mbox{.} in \cite{piomak,makpio,szczpio}.)
Second, in the asset closed games with incomplete information the player's profit as the logarithmic rate fulfils the first law of thermodynamics \cite{pioschro}.
What are the connections between the first and the second observations?
The considerations presented bellow try to answer this question. This problem concern
 the gamblers strategies based on Kelly criterion \cite{kelly,pioschro}.

\section{Quantitative transitivity property}
\label{}
Let us consider the transitivity property of the sequence $u:M\negthinspace\times\negthinspace M\rightarrow\mathbb{R}$ on quantities of goods.
In the absence of transaction costs (or, dually, when we neglect market scale effects) the relative prices (or the growth ratios in time) $u_{XY}$ fulfil
\begin{equation}
u_{AB} \odot u_{BC}\odot u_{CA}=1 \,, 
\label{pierw}
\end{equation}
where $\odot$ is any multiplication, $u_{XY}\in\mathbb{R}_+$,
$A,B,C,...$ are the symbols of the units of goods, or the moments of the market time.
Let us describe the above property $(\ref{pierw})$ in the logarithmic quantities that fulfil the transitivity law in
additive form that is convenient for statistical operations:
\begin{equation}
r_{AB} \oplus r_{BC}\oplus r_{CA}=0 \,, 
\label{drugie}
\end{equation}
where $r_{XY}:=\ln u_{XY}$. Further on we will call ``addition'' $\oplus$ the r-addition.

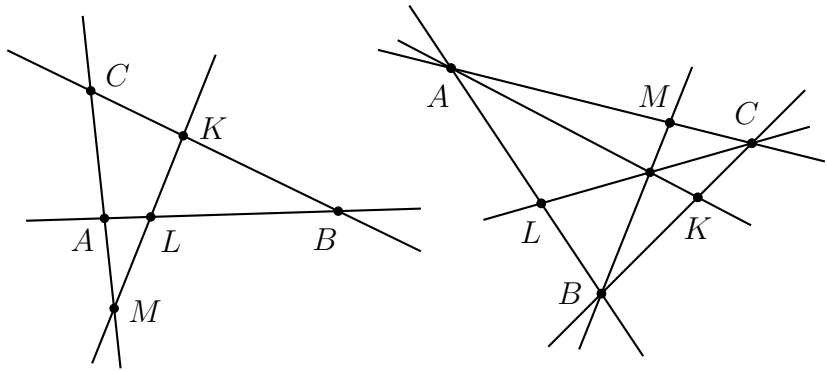
\begin{figure}[h]
\begin{center}
\begin{pspicture}(-1,-1)(10,4)
\pstGeonode[PointName=none,PointSymbol=none](.6,0){A1}(.5,0){B1}(.21,2.7){C1}(1.5,2.25){D1}
\pstGeonode[PointName=none,PointSymbol=none](0,2.8){E1}(.35,1){F1}(3.7,1){G1}(3.6,1.1){H1}\psset{linecolor=black,nodesep=-1}
\pstLineAB{A1}{D1}\pstLineAB{B1}{C1}
\psset{linecolor=black}
\pstLineAB{E1}{G1}
\pstLineAB{F1}{H1}
\psset{linecolor=black,linestyle=solid}
\pstInterLL[PosAngle=225]{B1}{C1}{F1}{H1}{A}
\pstInterLL[PosAngle=245]{E1}{G1}{F1}{H1}{B}
\pstInterLL[PosAngle=30]{B1}{C1}{E1}{G1}{C}
\pstInterLL[PosAngle=15]{A1}{D1}{E1}{G1}{K}
\pstInterLL[PosAngle=-50]{A1}{D1}{F1}{H1}{L}
\pstInterLL[PosAngle=-5]{A1}{D1}{C1}{B1}{M}
\psset{linecolor=black,linestyle=solid}
\pstGeonode[PosAngle=245](5,3){A}
\pstGeonode[PosAngle=180](7,0){B}
\pstGeonode[PosAngle=100](9,2){C}
\pstGeonode[PointName=none,PointSymbol=none](5.5,1){K1}(8,2.5){L1}
\pstInterLL[PosAngle=250]{K1}{C}{A}{B}{L}
\pstInterLL[PosAngle=120]{L1}{B}{A}{C}{M}
\psset{linecolor=black}
\pstInterLL[PointName=none]{L}{C}{M}{B}{O}
\psset{linecolor=black}
\pstInterLL[PosAngle=270]{A}{O}{B}{C}{K}
\psset{linecolor=black,nodesep=-1}
\pstLineAB{A}{B}\pstLineAB{B}{C}\pstLineAB{C}{A}
\psset{linecolor=black,,nodesep=-.8}
\pstLineAB{A}{K}
\pstLineAB{B}{M}
\pstLineAB{C}{L}
\psset{linecolor=black}
\pstGeonode[PointName=none](5,3){A}
\pstGeonode[PointName=none](7,0){B}
\pstGeonode[PointName=none](9,2){C}
\pstInterLL[PointName=none]{K1}{C}{A}{B}{L}
\pstInterLL[PointName=none]{L1}{B}{A}{C}{M}
\pstInterLL[PointName=none]{A}{O}{B}{C}{K}
\end{pspicture}
\end{center}
\label{figurka}
\caption{Mutually dual ancient nomographs (slide rules), respectively Menelaus' and Ceva's theorems: $\tfrac{p_{1AC}}{p_{2AC}}\,=\, \tfrac{p_{1AB}}{p_{2AB}}\,\tfrac{p_{1BC}}{p_{2BC}}$,
where:~$p_{1AB}=\lambda_{AB}|AL|$, $p_{2AB}=\lambda_{AB}|LB|$, $p_{1BC}=\lambda_{BC}|BK|$,
$p_{2BC}=\lambda_{BC}|KC|$, $p_{1AC}=\lambda_{AC}|AM|$, $p_{2AC}=\lambda_{AC}|MC|$, for~$\lambda_{AB},\lambda_{BC},\lambda_{AC}\in\mathbb{R}_+$.}
\end{figure}

In the ring of real numbers we can write Eq.~$(\ref{drugie})$ in the form of a condition,
that the value of some function  $f\negthinspace:\mathbb{R}^3\rightarrow\mathbb{R}$ is zero:
 $$
 f(r_{AB},r_{BC},r_{CA})=0\,.
 $$
 We require that this equation has a unique solution, that is symmetric with respect to all permutations
 of indexes $A,B,C$,  scale 
free and $r_{XX}=0$\,.
Then we can restrict generally this function $f$ to be of the following linear form (when we choose the natural units for $r_{XY}$):
$$
r_{AB}+r_{BC}+r_{CA} =0 
$$
for unbounded domain of $r_{XY}$ (that is well known case of logarithmic rate),  or
$$
r_{AB}+r_{BC}+r_{CA}+ r_{AB}\,r_{BC}\,r_{CA}=0\,\,\,\text{when}\,r_{AB}\in[-1,1]\,.
$$ 
After ordering, we have:
\begin{equation}
r_{AC}=r_{AB}\oplus r_{BC}:=\frac{ r_{AB}+r_{BC}}{1+r_{AB}\, r_{BC}}\,.
\label{stw}
\end{equation}
The above formula looks like the Einstein velocity addition law
of  the special relativity theory, expressed in the unit of the light
velocity (the maximal velocity $c=1$). Further considerations will be devoted to this very bounded case of ``rate''.
Substituting (probability Ansatz) $r_{XY}= p_{1XY}- p_{2XY}$, where $p_{jXY}\in[0,1]$ and $p_{1XY}+ p_{2XY}=1$ we obtain the r-addition rule $(\ref{stw})$ in the simplest form:
\begin{equation}
 \frac{p_{1AC}}{p_{2AC}}\,=\, \frac{p_{1AB}}{p_{2AB}}\,\frac{p_{1BC}}{p_{2BC}}\,,
 \label{baba}
 \end{equation}
  or (in the logarithms) in the form of ordinary addition on the group $(\mathbb{R},+)$.
 For this reason this logarithmic ``rate of rate'' $\tfrac{1}{2}\,\ln \tfrac{p_{1XY}}{p_{2XY}}$ is  convenient in calculations and is known as {\em rapidity}\/ in the special theory of relativity. Eq.~$(\ref{baba})$ is the unique projective invariant
known by ancient mathematicians as Menelaus' theorem (see Fig.~$\ref{figurka}$) \cite{grun}.
\section{Canonical model: two-horse race}
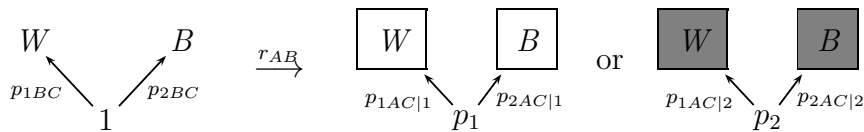
\begin{figure}[h]
\begin{center}
\begin{pspicture}(-6.5,0)(10,1.6)
\psset{linewidth=.7pt}
\fboxrule=2mm
\psset{labelsep=2pt}
\rput(-5,1){\rnode{A}{\em W}}
\rput(-3,1){\rnode{B}{\em B}}
\rput(-4,-.07){\rnode{C}{$1$}}
\ncline[nodesep=2pt]{->}{C}{A}
\Aput{$\scriptstyle{p_{1BC}}$}
\ncline[nodesep=2pt]{->}{C}{B}
\Bput{$\scriptstyle{p_{2BC}}$}
\rput(-1.7,.7){\rnode{A}{$\stackrel{r_{AB}}{\longrightarrow}$}}
\fboxrule=2\fboxrule
\rput(-.2,1){\rnode{A}{\frame{\colorbox[gray]{1}{\em W}}}}
\rput(1.6,1){\rnode{B}{\frame{\colorbox[gray]{1}{\em B}}}}
\rput(.8,-.07){\rnode{C}{$p_{1}$}}
\ncline[nodesep=2pt]{->}{C}{A}
\Aput{$\scriptstyle{p_{1AC|1}}$}
\ncline[nodesep=2pt]{->}{C}{B}
\Bput{$\scriptstyle{p_{2AC|1}}$}
\rput(2.7,.7){\rnode{A}{or}}
\rput(3.8,1){\rnode{A}{\frame{\colorbox[gray]{.5}{\em W}}}}
\rput(5.6,1){\rnode{B}{\frame{\colorbox[gray]{.5}{\em B}}}}
\rput(4.8,-.07){\rnode{C}{$p_{2}$}}
\ncline[nodesep=2pt]{->}{C}{A}
\Aput{$\scriptstyle{p_{1AC|2}}$}
\ncline[nodesep=2pt]{->}{C}{B}
\Bput{$\scriptstyle{p_{2AC|2}}$}
\end{pspicture}
\end{center}
\label{figugurka}
\caption{F-gambler's knowledge cracks space of events.}
\end{figure}

The bounded domain of the logarithmic rate $r_{XY}$ appears not only in the special theory of relativity. We can observe this case in a lot of simple stochastic models. 
Let us assume that in the bookmaker game the gambler (say: the forecast-gambler, f-gambler) predicts future results
of, for simplicity, two-horse races better than the average gambler.
Let $p_{1AB}$ be the probability of correctness of her/his forecast, and let $p_{2AB}$ be
the probability of her/his mistake. F-gambler's knowledge $r_{AB}$ decomposes two element space of events (the race results)
 in four element space, see Fig.~\ref{figugurka}. 

Indexes $1$, $2$, $1BC$, $2BC$, $1(2)AC|1$, denote respectively the events:   ``{\em the f-gambler places a bet on the white horse (W)}\/'', ``{\em the f-gambler places a bet on the black horse (B)}\/'', ``{\em the white horse win}\/'', ``{\em the black horse win}\/'',  and ``{\em good (bad) guessing when the white horse win}\/''.  The indexes to  the left of the vertical bar $|$ represent the events and the expressions to the right of the vertical bar represent the conditions. $r_{AB}=p_{1AB}-p_{2AB}$.
We assume that events $AB$, $BC$ are stochastically independent.

Let us describe (c.f.~\cite{pioschro}) the sum of wagers from all the gamblers of the bet on event $j$
by $I\negthinspace N_j$, and the percentage share of f-gambler's capital in four cases of
bookmakers' bets by $l_{jk}$ (or equivalently by $l_{j|k}$), $L=(l_{kj})$. F-gambler's profit in a concrete horse race is equal to  the projective invariant logarithmic rate \cite{pioschro}:
$$
z_{k|j}=z_{kj}:= \ln(1 + \tfrac{IN_{3-k}}{IN_k}
\, l_{kj}-l_{(3-k)j})\,.
$$
And let us form the matrix $P$ of probabilities of all four elementary events:
$$
P=(p_{kj}):=
\left( \begin{array}{cc}p_{1AB}p_{1BC}&p_{2AB} p_{1BC}\\ p_{2AB}p_{2BC}&p_{1AB} p_{2BC}\end{array} \right) .
$$
Then the f-gambler expected profit is equal to:  
$$
E_Z(L):=\sum_{k,j=1}^2 p_{kj}z_{kj}=\sum_{k,j=1}^2 p_{kj}\ln(1 + \tfrac{IN_{3-k}}{IN_k}
\, l_{kj}-l_{(3-k)j})\,,
$$
where $Z:=(z_{kj})$. The rational f-gambler bets the stakes  $\bar{L}=(\bar{l}_{kj})$ such that
her/his expected profit is the maximal one:
$$ 
E_Z(\bar{L}):=\max_{l_{11},l_{12},l_{21},l_{22}} \{E_Z(L)\}\,.
$$
Now, rationally, we find the global maximum of the function $E_Z(L)$.

\section{Maximizing profit strategy}
After standard differential calculations we obtain that the f-gambler's maximal profit is given by:
\begin{equation}
E_Z(\bar{L})= D_Z-S_{AC}\,.
\label{zysk}
\end{equation}
The first component of the maximal profit in $(\ref{zysk})$ is the seer's profit
i.e\mbox{.} the profit on unpopularity of the winning bet:
$$D_Z=-\sum_{j=1,2}
p_{jBC}\ln \tfrac{I\negthinspace N_j}{I\negthinspace
N_1+I\negthinspace N_2}\, ,
$$ 
and the second denotes the (minus) Boltzmann/Shannon entropy
$-S_{AC}$ of the forecast:
$$
S_{AC}:= - p_1 \sum_{j=1}^2 p_{jAC|1}\ln p_{jAC|1} - p_2 \sum_{j=1}^2 p_{jAC|2}\ln p_{jAC|2}\,. 
$$ 
The profit difference $-S_{AC}-\sum_j p_{jBC}\ln p_{jBC}$ between the rational forecast and the seer's forecast ($r_{AB}=0$)
does not depend on the sums $I\negthinspace N_j$ of wagers from all the gamblers of the bet.

The two families 
$(\bar{l}_{1j},\bar{l}_{2j})\negthinspace\in\negthinspace\mathbb{R}^2$, $j=1,2$, of the  solutions--strategies of extremal problem are described by the following two straight line equations:
\begin{equation}
(\bar{l}_{jj}-p_{1AC|j})\,I\negthinspace N_{3-j}=(\bar{l}_{(3-j)j}-p_{2AC|j})\,I\negthinspace N_j,\,\,\,j=1,2.
\label{oola}
\end{equation}
In the absence of short positions (a typical restriction on the bets 
$\bar{l}_{1j},\bar{l}_{2j}\geq 0$) we assume that the rational f-gambler 
diversificates the risk in such a way that she/he bets only possible a 
minimal part of her/his resources.  From all the strategies $(\ref{oola})$ we choose 
the optimal one:
$$
\begin{array}{rl}
\text{\tt\large {if}}&\,\,p_{1AC|j}I\negthinspace N_{3-j}>p_{2AC|j}I\negthinspace N_j\\
\text{\tt\large {then}}&\,\, \bar{l}_{jj}^\ast=p_{1AC|j}-p_{2AC|j}\tfrac{I\negthinspace N_j}{I\negthinspace N_{3-j}}\,,\,\,\,\,\bar{l}_{(3-j)j}^\ast=0
\\
\text{\tt\large {else}}&\,\,
\bar{l}_{(3-j)j}^\ast=p_{2AC|j}-p_{1AC|j}\tfrac{I\negthinspace N_{3-j}}{I\negthinspace N_{j}}\,,\,\,\,\,\bar{l}_{jj}^\ast=0\,.
\end{array}
$$

If we do not have the information about proportion of sums of wagers $\tfrac{IN_1}{IN_2}$
then we use famous {\em Laplace's Principle of Indifference}\/ $(IN_1=IN_2)$, and the optimal stakes are
(in the form of the step function with jump at zero, $[r]_+:=\max\{r,0\}$):
\begin{equation}
 \bar{L}^\ast=
\left( 
\begin{array}{rr
}  
{[r_{AC|1}]_+} & {[-r_{AC|2}]_+}\\
{[-r_{AC|1}]_+} & {[r_{AC|2}]_+}
\end{array}
\right)\,.
\label{statnie}
\end{equation}
We have four cases of play: normal game (max-strategy is diagonal), mis\`ere game (antidiagonal) and two mixed normal-mis\`ere games (horizontal).

\section{Conclusion}
The formal description of the bookmaker bets with majority
of branches of events might be created hierarchically as the binary
tree with the leafs -- elementary events, e.g\mbox{.} by analogy
to the construction of tree-shaped key to
compressing/decompressing Huffman code \cite{cormen}. It follows
that our binary bet is {\em universal}\/, i.e\mbox{.} many kinds of financial decisions
we can describe as the systems based on a hierarchy of formal binary bets.
The bet model presented in \cite{pioschro} and above, as a projective covariant counterpart for basic in financial economics instruments there are that Arrow-Debreu securities \cite{arrow,fortnow} has interesting econophysics
interpretation in the fields of thermodynamics and special theory of relativity. 
In the absence of information about the wagers decomposition the pair  $(r_{AC|1},r_{AC|2})$ of ``relativity-stochastic'' rates in Eq\mbox{.} $(\ref{statnie})$ fully determines the optimal stakes of f-gambler.
Within presented motivation, in postmodern manner,  such a rule we can call {\em the Menelaus-Kelly criterion}\/ for bets \cite{kelly}. 




\begin{thebibliography}{00}
\bibitem[1]{piosla} E.~W.~Piotrowski, J.~S\l adkowski, {\em Geometry of Financial Markets -- Towards Information Theory Model of Markets}\/, Physica A, 382 (2007) 228-234.
\bibitem[2]{piomak}  E.~W.~Piotrowski, M.~Makowski, {\em Cat's Dillema -- Transitivity vs.~Intransitivity}\/, Fluc.~Noise Lett\mbox{.} 5 (2005) L85-L95.
\bibitem[3]{makpio} M.~Makowski, E.~W.~Piotrowski, {\em Quantum Cat's Dilemma: an example of intransitivity in  a quantum game}\/, Phys.~Lett.~A, 355 (2006) 250-254.
\bibitem[4]{szczpio} A.~Szczypi{\'n}ska, E.~W.~Piotrowski, {\em Projective Market Model Approach to AHP Decision--Making}\/, APFA-6 Conference, Physica A, in press.
\bibitem[5]{pioschro} E.~W.~Piotrowski, M.~Schroeder, {\em Kelly Criterion Revisited: Optimal Bets}\/, Eur.~Phys.~J.~B, 57 (2007) 201-203.
\bibitem[6]{kelly} J.~L.~Kelly, Jr., {\em A New Interpretation of Information Rate}\/,
 Bell System Technical Journal 6 (1956) 917-926; http://www.arbtrading.com/kelly.htm.
\bibitem[7]{grun} B.~Gr\"unbaum, G.~C.~Shephard,  {\em Ceva, Menelaus, and Selftransversality}\/, Geometriae Dedicata,
65 (1997) 179-192.
\bibitem[8]{cormen} T.~H.~Cormen, C.~E.~Leiserson, R.~L.~Rivest, C.~Stein. {\em Introduction to Algorithms}\/,  MIT Press, Cambridge MA and McGraw--Hill, New York, 2001.
\bibitem[9]{arrow} K.~J.~Arrow, {\em Essays in the Theory of Risk-Bearing}\/, North-Holland Pub.~Co., Amsterdam, 1971.
\bibitem[10]{fortnow} L.~Fortnow, J.~Kilian, D.~M.~Pennock, M.~P.~Wellman, {\em Betting Boolean-style: a framework for trading in securities based on logical formulas}\/, Decision Support Systems 39(1) (2005) 87-104.


\end{thebibliography}
\end{document}